# Transition Matrix Analysis: Analyzing Students' Use of Cognitive Resources in Physics

Tianlong Zu,[1] and N. Sanjay Rebello[2,3]
[1]*Department of Physics and Astronomy, Northwestern University, 2145 Sheridan Road, Evanston, IL, USA 60208*
[2]*Department of Physics and Astronomy, Purdue University, 525 Northwestern Ave., West Lafayette, IN 47907*
[3]*Department of Curriculum and Instruction, Purdue University, 100 N. University St., West Lafayette, IN 47907*

**Abstract**: Conceptual surveys of multiple-choice format have been developed to test the effect of pedagogical interventions on students' understanding of physics knowledge. Predominantly, they are administered in pre- and posttest settings and analyzed to obtain a performance gain. However, focusing on the correct answers to each question alone ignores the incorrect options which could inform us the stability and coherency of the knowledge structure of students. According to the resource model framework, each specific answer from students could reflect a specific cognitive resource being activated and implemented in the context at hand. Consequently, conceptual surveys could demonstrate how instruction could affect the activation of different cognitive resources of students in a variety of contexts when administered before and after instruction. Guided by the resources framework, we propose a transition matrix analysis to analyze data collected through conceptual surveys to investigate how instruction affects the consistency of cognitive resources activation by students. To provide proof of concept, we demonstrated how to utilize this method by analyzing students' responses to a subset of questions from the DIRECT survey in a classroom study.

## Introduction

Many physics educators embrace a constructivist approach teaching and learning which requires us to know what prior knowledge students bring to our classrooms, how and when they activate that knowledge in various contexts, and how students respond to different instructional methods [1]. To this end, research validated conceptual surveys such as Force Concept Inventory (FCI) [2], Force and Motion Conceptual Evaluation (FMCE) [3], Determining and Interpreting Resistive Electric Circuit Concepts Test (DIRECT) [4], and others (for review, see [5]) have been widely used to assess students' conceptual understanding, both pre- and post- instruction. These conceptual surveys were developed based on extensive qualitative research that has documented students' difficulties and common incorrect answers to conceptual questions [5]. During the time these conceptual surveys were developed and since, some researchers held the view that the naïve knowledge brought to classroom by novice learners of physics could be described as "misconceptions," "alternative conceptions," or "naïve theories" which were relatively coherent stable cognitive structures [6,7,8,9]. Other researchers held the view that students' naïve knowledge was fragile and spontaneous [10,11]. Research suggests the knowledge of novice students of physics are fragmented, unstable and small grain-sized [12,13,14]. As per this view, knowledge is a set of small, intuitive ideas (resources) that students may apply either consistently or inconsistently depending on the context [15]. The resources framework encourages educators to identify the useful "prior knowledge" of students on which instructors can build new knowledge [11,16]. In this work, we propose that research validated conceptual surveys can be used to extract resources students use in different contexts.

Researchers have studied different types of resources, such as epistemological resources [17,18,19], and conceptual resources [13]. In this work, we focus only on conceptual resources. We regard cognitive resources as very basic units of knowledge piece students activate in a specific situation to form a concrete answer. One underlying assumption of this work is that each choice to a multiple-choice question on a conceptual survey reflects a specific cognitive resource of students. This means the resources used by different students are the same if they select the same choice to a question (if not due to random selection). The resource framework can also provide an explanation for formation of misconception, i.e. the relative stability of cognitive structure can be explained as associated cognitive resources tended to be activated together.

In this work, we introduce a transition matrix analysis method and provide a proof of concept by analyzing data collected via research validated conceptual surveys to extract and examine cognitive resources activation by students. This study could also contribute to the increasing interests in physics education community using mathematical modelling to describe the learning processes in recent years [20,21,22,23,24].

## Transition Matrix Analysis

To adopt transition matrix analysis, a subset of questions called resources-equivalent questions are identified on research validated conceptual survey. The so-called resources-equivalent questions are a set of questions that should be answered with the same underlying physics principle but designed in different contexts. In addition, the whole set of choice options to each question should correspond to the same set of cognitive resources. It would indicate that students have activated different cognitive resources in different contexts if they do not answer the resources-equivalent questions consistently.

Inspired by the probabilistic view of student thinking in the work of Bao and Redish [20], we propose the level of inconsistency of cognitive resources activation by a single student can be represented by a state vector. Imagine there are $w$ cognitive resources that students use for solving $m$ questions which form a set of resources-equivalent questions. Define $\vec{Q}^k$ as the $k$th student's probability distribution vector measured with the $m$ questions. Then we can write:

$$\vec{\boldsymbol{Q}}^k = \begin{pmatrix} \rho_1^k \\ \rho_2^k \\ \vdots \\ \rho_w^k \end{pmatrix},$$

where $\rho_i = n_i/m$ represents the probability of selecting $i$th cognitive resource and $n_i$ is the number of questions answered using cognitive resource $i$. Among these $w$ cognitive resources, there is only one that is applied in the scientifically correct way which is represented by one element of the vector. All the elements should add up to unity since $\sum_{i=1}^{w} n_i = m$.

An important objective of various pedagogical interventions is to track how students utilize different cognitive resources. For a set of resources-equivalent questions, two probability distribution vectors are assigned to a student for the pre- and posttest. We propose these two vectors are connected by a transition matrix which allows researchers to see how students change the cognitive resources activation from pretest to posttest. It is more informative than just comparing the two initial and final state vectors because it can show if the intervention is effective in activating the cognitive resources of students across various contexts. To serve this purpose, for student $k$, the two probability vectors are connected by a matrix:

$$\boldsymbol{Q}_f^k = \boldsymbol{T}\boldsymbol{Q}_i^k,$$

where $\boldsymbol{T}$ represents the transition matrix, $\boldsymbol{Q}_i^k$ represents the initial state vector of the probability distribution, and $\boldsymbol{Q}_f^k$ represents the final state vector of the probability distribution. In matrix format, the above equation is expressed as:

$$\begin{pmatrix} \rho_{1f}^k \\ \rho_{2f}^k \\ \vdots \\ \rho_{wf}^k \end{pmatrix} = \begin{pmatrix} T_{11}^k & T_{12}^k & \cdots & T_{1w}^k \\ T_{21}^k & T_{22}^k & \cdots & T_{2w}^k \\ \vdots & \vdots & \ddots & \vdots \\ T_{w1}^k & T_{w2}^k & \cdots & T_{ww}^k \end{pmatrix} \begin{pmatrix} \rho_{1i}^k \\ \rho_{2i}^k \\ \vdots \\ \rho_{wi}^k \end{pmatrix}$$

In the above matrix, an element $T_{ij}^k$ represents how a student changes the cognitive resources used from pretest to posttest:

$$T_{ij}^k = \frac{n_{ij}^k}{n_j^k},$$

where $n_{ij}^k$ represents the number of questions answered with cognitive resource $j$ on pretest but answered with cognitive resource $i$ on the posttest by student $k$. And $n_j^k$ represents the number of questions answered with cognitive resource $j$ on the pretest. This matrix describes the redistribution of the probability of different cognitive resources. Note: if $n_j^k$ is zero on the pretest, then all elements in the $j$th column of the matrix will be zero.

We could expand these expressions to some subgroups or the whole class. Imagine a class of $N$ students, a transition matrix can be identified for each student and a transition matrix for, for example, the whole class can be obtained by calculating the average of all students' individual matrices:

$$\begin{pmatrix} T_{11} & T_{12} & \cdots & T_{1w} \\ T_{21} & T_{22} & \cdots & T_{2w} \\ \vdots & \vdots & \ddots & \vdots \\ T_{w1} & T_{w2} & \cdots & T_{ww} \end{pmatrix} = \frac{1}{N}\sum_k \begin{pmatrix} T_{11}^k & T_{12}^k & \cdots & T_{1w}^k \\ T_{21}^k & T_{22}^k & \cdots & T_{2w}^k \\ \vdots & \vdots & \ddots & \vdots \\ T_{w1}^k & T_{w2}^k & \cdots & T_{ww}^k \end{pmatrix}.$$

The transition matrix for the whole class can inform how the students on average change the cognitive resources for answering a set of resources-equivalent questions identified on any conceptual tests. In this case, $T_{ij}$ represents the average percentage of questions answered with resource $j$ on the pretest that are answered with resource $i$ on the posttest[1].

### Transition Matrix Analysis of DIRECT data

We applied this transition matrix analysis to a set of data collected from students enrolled in a conceptual physics class for elementary education majors at a large land grant mid-western university. The material used in this study is DIRECT developed by Engelhardt and Beichner [4] to test students' understanding of electric circuits. The version of DIRECT used in this study includes 29 MC questions, each with 5 choices. DIRECT was administered to N=135 students as a pretest and posttest of an instructional unit on electric circuits. At both times, students first worked individually and then in a group of three or four on the survey (see Fig. 1). Students were encouraged to provide individual responses during the group session. They were not required to reach consensus, and no TA intervention occurred at any time. In fact, students did not always have unanimous answers to all questions during group work. There is a 7-week long instructional period between the pretest and posttest. During this time, students attended one 50-minute lecture and a 170-minute lab per week without any other extra-curricular activities. The textbook was *Conceptual Physics* of Hewitt [25].

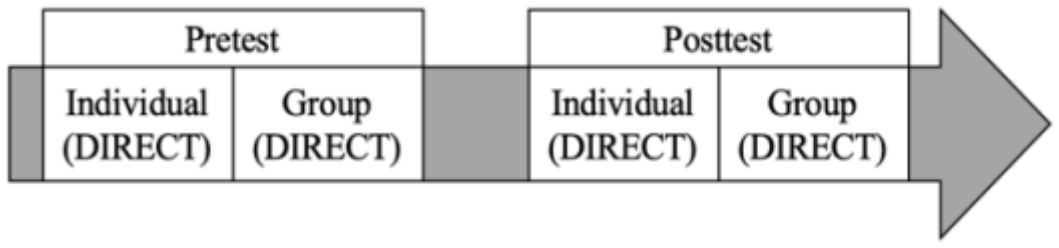

Figure 1: Procedure of this study.

Prior research has documented general student understanding with respect to direct current circuits questions [4]. For example, student would answer with the following reasoning: i) current is consumed within the circuit, ii) battery is a source of constant current, iii) students using sequential reasoning believe current is only influenced by each element when the current encounters that element [13,26], iv) students using local reasoning think that the current is always divided into equal parts whenever a junction is encountered, v) students using superposition reasoning believe that two batteries make a bulb shine twice as bright as one battery regardless of the configuration (for more, see [4]).

It has been shown that students do not use these different ways of reasoning consistently [27]. Rather, selection of different ways of reasoning depends on the cues embedded in a specific question, such that cognitive resources activation becomes context dependent [21].

---

[1] Note that different students may have different $\rho_{ji}^k$, so we do not expect the class average matrix elements in the $j$th collum to add to 1.

A set of resources-equivalent questions can be identified on the DIRECT test (Q3, Q7, Q12)[2] (click the footnote for details). The three questions examine students' understanding of the concept: "*power of direct current circuits*". To have diagnostic function, different choices of these questions include distracters related to the same set of unproductive cognitive resources. We have made a connection between different choices of each question to the set of cognitive resources. The detailed mapping between the four cognitive resources and choices to each question can be found in table 1.

- $S$: *scientifically productive cognitive resource*.
- $R_1$: the resource of constant source applied to *battery as a constant current source which says battery supplies same amount of current to each circuit regardless of the circuit's arrangement.*
- $R_2$: *the resource of more cause leads to more effect rising to battery superposition, which says bulbs shine brighter with two batteries regardless of arrangement* [15].
- $N$ or $Null$: *the choices that corresponding to cognitive resources that are used by very few students*. (It is necessary to have this option to maintain the completeness of the state vector).

Let's take question 3 as an example. If students select B as their answer, the most probable reasoning might be: "*battery as a constant current source"* [4,27], such that they would argue the current from the two branches in circuit 2 will add up when they meet in a joint point and then provide more power compared to circuit 1&3 since circuits 1&3 have only one branch respectively. When students select E as their answer, they might have argued with another common reasoning: "*battery superposition*" [4,27]. Since both circuits 2&3 both have two batteries, they should provide more power than circuit 1 which have only one battery regardless of the arrangement of the batteries in circuits 2&3.

Table 1: Association between resources and the choices of the three questions on DIRECT.

| Question | S | R1 | R2 | Null |
|---|---|---|---|---|
| Q3 | *C* | *B* | *E* | *A, D* |
| Q7 | *B* | *D* | *E* | *A, C* |
| Q12 | *D* | *E* | *A* | *B, C* |

Note that table 1 was a first attempt using the resources framework since the authors of this paper are aware that DIRECT was developed following the misconception framework at the time. We acknowledge it is a limitation of this study and hence urge the readers to treat it as an assumption.

Before we go into the details of transition matrix analysis, we report the overall performance (of all 29 questions) of students on the DIRECT test. A paired samples t-test revealed a significant difference between students' performance of individual pretest (mean = 7.03, S.D. = 2.27) and that of group pretest (mean = 8.15, S.D. = 2.48), $t(134) = -4.809$, $p < .001$. Similarly, a paired samples t-test revealed a significant difference between the performance of individual posttest (mean = 9.62, S.D. = 2.99) and group posttest ((mean = 11.87, S.D. = 2.80), $t(134) = -9.071$, $p < .001$.

To investigate the impact of group work on the selection of cognitive resources, we applied transition matrix analysis to the data collected from the individual session and the group session of the pretest (Results can be found in Table 2) and the posttest (Results can be found in Table 3). Notice that the Null cognitive resource option was the least selected indicating these resources were least activated and applied by students while answering the three questions.

Tables 2 & 3 demonstrate similar result patterns which showed that many students changed from productive cognitive resource $S$ in the individual session to unproductive cognitive resources ($R_1$ and $R_2$) in the group session as indicated by $T_{R_2R_1}$, and $T_{R_2S}$ in the class transition matrices. We found that the class maintained many of their original selection after group work as indicated by the diagonal elements ($T_{SS}$, $T_{R_1R_1}$, and $T_{R_2R_2}$). The last three elements in the first row of the class transition matrix indicate the degree to which students change their activation of unproductive resources to the productive one. The results indicate that many students retain their use of unproductive cognitive resources after instruction, and the most unproductive one is $R_2$. Students do not typically switch between unproductive cognitive resources ($T_{R_1R_2}$ and $T_{R_2R_1}$). Some students showed switching from productive resources to unproductive resources ($T_{R_1S}$ and $T_{R_2S}$). In summary, the TMA results indicate that even though students significantly improved their overall performance on DIRECT (total class average), students do not typically have solid understanding of the concept of power of direct current circuits. This important information would be ignored if we only report the overall gain in performance.

2 https://drive.google.com/file/d/1fN4n4Yfn1bLdX4QkmOAopuLerz2RSmr6/view?usp=sharing

Table 2: TMA results averaged over all participants. (pretest IN & pretest G).

| Pretest IN class average state vector | Pretest G class average state vector |
|---|---|
| $\begin{pmatrix}\rho_S\\ \rho_{R_1}\\ \rho_{R_2}\\ \rho_N\end{pmatrix}=\begin{pmatrix}.293\pm.023\\ .213\pm.022\\ .316\pm.025\\ .178\pm.021\end{pmatrix}$ | $\begin{pmatrix}\rho_S\\ \rho_{R_1}\\ \rho_{R_2}\\ \rho_N\end{pmatrix}=\begin{pmatrix}.203\pm.024\\ .346\pm.032\\ .356\pm.027\\ .095\pm.014\end{pmatrix}$ |

$$\begin{pmatrix}T_{SS} & T_{SR_1} & T_{SR_2} & T_{SN}\\ T_{R_1S} & T_{R_1R_1} & T_{R_1R_2} & T_{R_1N}\\ T_{R_2S} & T_{R_2R_1} & T_{R_2R_2} & T_{R_2N}\\ T_{NS} & T_{NR_1} & T_{NR_2} & T_{NN}\end{pmatrix} = \begin{pmatrix}.231\pm.036 & .079\pm.023 & .073\pm.018 & .044\pm.016\\ .253\pm.037 & .274\pm.038 & .119\pm.026 & .095\pm.024\\ .145\pm.030 & .104\pm.024 & .358\pm.039 & .187\pm.032\\ .010\pm.008 & .024\pm.012 & .089\pm.022 & .088\pm.022\end{pmatrix}$$

Table 3: TMA results averaged over all participants. (posttest IN & posttest G).

| Posttest IN class average state vector | Posttest G class average state vector |
|---|---|
| $\begin{pmatrix}\rho_S\\ \rho_{R_1}\\ \rho_{R_2}\\ \rho_N\end{pmatrix}=\begin{pmatrix}.190\pm.023\\ .341\pm.027\\ .373\pm.025\\ .095\pm.014\end{pmatrix}$ | $\begin{pmatrix}\rho_S\\ \rho_{R_1}\\ \rho_{R_2}\\ \rho_N\end{pmatrix}=\begin{pmatrix}.128\pm.022\\ .411\pm.034\\ .414\pm.029\\ .048\pm.010\end{pmatrix}$ |

$$\begin{pmatrix}T_{SS} & T_{SR_1} & T_{SR_2} & T_{SN}\\ T_{R_1S} & T_{R_1R_1} & T_{R_1R_2} & T_{R_1N}\\ T_{R_2S} & T_{R_2R_1} & T_{R_2R_2} & T_{R_2N}\\ T_{NS} & T_{NR_1} & T_{NR_2} & T_{NN}\end{pmatrix} = \begin{pmatrix}.122\pm.027 & .038\pm.017 & .076\pm.022 & .060\pm.021\\ .134\pm.029 & .445\pm.042 & .173\pm.032 & .060\pm.021\\ .144\pm.029 & .153\pm.030 & .481\pm.041 & .102\pm.026\\ .014\pm.009 & .004\pm.004 & .036\pm.012 & .041\pm.017\end{pmatrix}$$

Table 4: Samples t-test results.

| Matrix element | $t(132)$ | $p$ |
|---|---|---|
| $T_{SS}$ | 2.613 | .010 |
| $T_{R_1S}$ | 2.466 | .015 |
| $T_{R_1R_1}$ | -3.060 | .003 |
| $T_{R_2R_2}$ | -2.619 | .010 |
| $T_{R_3N}$ | 2.101 | .038 |
| $T_{NR_2}$ | 2.091 | .038 |

We also conducted paired samples t-tests to compare all the elements in the two class matrices in Table 4 (note only results of significant differences were included). We found that, compared to pretest, students in the posttest retained significantly less productive cognitive resource in the group session ($T_{SS}$). Students retain their unproductive cognitive resources in the group session on significantly more questions in the posttest compared to pretest ($T_{R_1R_1}$ and $T_{R_2R_2}$). These results provide evidence for the relative stability of the activation of unproductive cognitive resources.

## General Discussion

In this work, we demonstrated a new approach guided by the resources framework to analyzing data collected through conceptual surveys. This method is particularly useful if these surveys consist of MC questions that can be treated as resources-equivalent questions.

To provide proof of concept, we identified a set of resources-equivalent questions from DIRECT. We demonstrated the process of transition matrix analysis by tracking how students applied cognitive resources in response to group work in both pre- and posttest. We found that students retain the two unproductive cognitive resources and some even switched from the productive resource to the unproductive ones during group work. This transition is worse in the posttest than in the pretest which should alert instructors to spend more time addressing the two unproductive cognitive resources in class.

Since DIRECT was developed following the misconceptions framework, it is challenging to explicitly identify the underlying cognitive resources associated with each response option. In this work, we made an initial attempt to establish a mapping between potential cognitive resources and students' responses to the three questions examined. Another underlying assumption of this study is that students have developed a certain degree of stability in their cognitive resources, such that different students selecting the same response option are likely activating similar cognitive resources. Future research is needed to further examine and validate these assumptions through methods (such as interview) that can more directly probe students' underlying reasoning processes.

Nonetheless, TMA provides a new perspective for examining how students activate different cognitive resources when responding to resource-equivalent questions from pretest to posttest. If this line of research proves fruitful, it would motivate subject-matter experts to revisit existing conceptual assessments and identify the possible sets of cognitive resources underlying students' response choices for each question. Furthermore, it would encourage researchers to intentionally design resource-equivalent questions when developing future conceptual surveys.

## Acknowledgements

This work is supported in part by the U.S. National Science Foundation grant 1348857 and 211138. Opinions expressed are of the authors and not necessarily of the Foundation.